# Developments Toward a 250-nm, Fully Planarized Fabrication Process With Ten Superconducting Layers And Self-Shunted Josephson Junctions

Sergey K. Tolpygo, Vladimir Bolkhovsky, Ravi Rastogi, Scott Zarr, Alexandra L. Day, Terence J. Weir, Alex Wynn, and Leonard M. Johnson

Quantum Information and Integrated Nanosystems Group
Lincoln Laboratory, Massachusetts Institute of Technology
Lexington, MA, USA 0242
sergey.tolpygo@ll.mit.edu

*Abstract*— We are developing a superconductor electronics fabrication process with up to nine planarized superconducting layers, stackable stud vias, self-shunted Nb/AlO$_x$-Al/Nb Josephson junctions, and one layer of MoN$_x$ kinetic inductors. The minimum feature size of resistors and inductors in the process is 250 nm. We present data on the mutual inductance of Nb stripline and microstrip inductors with linewidth and spacing from 250 nm to 1 µm made on the same or adjacent Nb layers, as well as the data on the linewidth and resistance uniformity.

## I. INTRODUCTION

Recent developments in the fabrication technology for superconductor electronics at MIT Lincoln Laboratory (MIT LL) [1],[2] have enabled very large scale integration (VLSI) of Single Flux Quantum (SFQ) circuits. For instance, ac-biased shift registers [3] with 202 kbit and nearly $10^6$ Josephson junctions (JJs) [4] have been demonstrated. Despite this progress, the main challenge for superconductor electronics has been scalability [5], i.e., the ability to progressively increase the number of logic cells in superconductor integrated circuits. These challenges arise because magnetic flux information encoding requires larger-area logic gates than their charge-based semiconductor counterparts.

The current MIT LL technology node SFQ5ee [2] has eight Nb layers, one layer of MoN$_x$ kinetic inductors, and utilizes resistively shunted Nb/AlO$_x$-Al/Nb Josephson junctions (RSJs) with the Josephson critical current density $J_c = 0.1$ mA/µm$^2$. The minimum feature size, $F$, is 350 nm. The area occupied by a circular junction in the plane of its base electrode or its top wiring layer (layers M5 and M6 in our process, respectively), is $\pi(r+F)^2$, where $r$ is the JJ counter electrode radius, $r = (I_c/\pi J_c)^{1/2}$ and $F$ is the minimum surround. A shunt resistor area is $l \cdot F + 3A_{via}$, where $l$ is the resistor length, $A_{via}$ is the via area; and three vias are required to connect a shunt resistor to a JJ. The resistor length of an $F$-wide resistor $R_s$ is $l = (R_s/R_{sq})F$, where $R_{sq}$ is the sheet resistance. The SFQ5ee process node offers two values of $R_{sq}$: 2 and 6 Ω/sq. Because each via needs to be covered by Nb wiring with surround $F$ on all sides, the minimum area of one via in the M6 plane is $A_{via} = 9F^2$. The total shunt resistor area is $A_R = (R_s/R_{sq})F^2 + 27F^2$.

The minimum area occupied by an RSJ is the sum of areas of the JJ, the shunt resistor, and the vias and is given by

$$A_{RSJ} = \pi[(I_c/\pi J_c)^{1/2} + F]^2 + [V_p/(I_c R_{sq}) + 27]F^2, \quad (1)$$

where $V_p$ is the process characteristic voltage. Equation (1) is a simplified version of equation (4a) in [5]. The maximum density of RSJs can be estimated as $n_{RSJ} = k/A_{RSJ}$, where $k \sim 0.5$ is the area fill-factor which takes into account that JJs need to be connected to inductors on various circuit layers and to other JJs.

The maximum density of RSJs in superconducting circuits as a function of the junction critical current is shown in Fig. 1 for different values of $F$ and $J_c = 0.1$ mA/µm$^2$ as in the SFQ5ee process node. For a comparison, we also show the RSJ maximum density for our newest process node, SFQ5hs, that has $J_c = 0.2$ mA/µm$^2$ but is identical in all other respects to the SFQ5ee node [6]. For 0.2-mA/µm$^2$ JJs, $V_p$ in (1) is about 0.86 mV [6]. The typical range of critical currents of junctions used in SCE digital circuits is from about 0.1 mA to 0.2 mA. It can be seen from Fig. 1 that the maximum density of RSJs can reach about $10^7$ cm$^{-2}$ if all vias and resistors can be fabricated with the feature sizes of 0.25 µm.

Large RSJ area (1) is currently the main limitation to increasing the integration scale of superconductor electronics. The area of SFQ cell inductors poses a similar scalability constraint because each JJ is typically connected to one inductor. Circuit density is maximized when inductor density matches the junction density. The typical cell inductance is $L = \beta_L \Phi_0/2\pi I_c$, where $\beta_L$ is the cell design parameter, typically in the range from 2 to 3. The area of an $F$-wide stripline inductor with spacing $F$ and two vias can be estimated as [5]

$$A_L = 2F(L/\ell) + 18F^2, \quad (2)$$

This research is based upon work supported by the Office of the Director of National Intelligence, Intelligence Advanced Research Projects Activity, via Air Force Contract FA872105C0002. The views and conclusions contained herein are those of the authors and should not be interpreted as necessarily representing the official policies or endorsements of the U.S. Government.

where $\ell$ is width-dependent stripline inductance per unit length. The latter was measured in [7] for multiple stripline combinations in our process. At $F = 0.25$ μm, the typical value is $\ell \approx 0.7$ pH/μm. The density of inductors is $n_L = m_L k/A_L$, where $m_L$ is the number of independent layers of inductors, $k$ is the area fill factor. Fig. 1 also shows the maximum density of stripline inductors $n_L$ estimated using (2) at $F = 0.25$ μm and two independent layers of inductors.

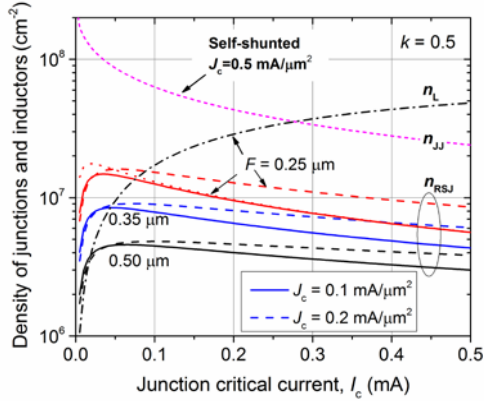

Fig. 1. The maximum density of resistively shunted junctions, $n_{RSJ}$ at the minimum feature sizes of 0.25, 0.35, and 0.50 μm, from top to bottom, and the critical current density of 0.1 and 0.2 mA/μm$^2$ corresponding to the solid and dashed curves, respectively. The short dash shows the maximum density of the self-shunted JJs, $n_{JJ}$ with $J_c = 0.5$ mA/μm$^2$. The dashed-dotted curve shows the maximum density of stripline inductors, assuming two independent layers $m_L = 2$ of inductors with the linewidth and spacing $F = 0.25$ μm, $\beta_L = 2$, $\ell = 0.7$ pH/μm, and the area fill factor $k = 0.5$ on each layer.

These projections show that achieving $F = 0.25$ μm for the cell inductors and shunt resistors would match the RSJ and inductor densities at about $10^7$ cm$^{-2}$, thus enabling SFQ circuits with slightly over $10^6$ logic gates per cm$^2$. The first goal of the work presented below is to realize this minimum feature size.

The JJ density can be further increased by removing the external shunts and transitioning to self-shunted JJs with higher $J_c$, as shown by the short-dash curve in Fig. 1. The second goal of this work is to integrate the self-shunted JJs into the process.

However, the achievable inductor density at $F = 0.25$ μm is insufficient to match the density of self-shunted JJs with $I_c$ in the typical range from about 0.05 mA to 0.2 mA and make a sufficient impact on the overall circuit density. This problem can be solved by transitioning from magnetic (geometrical) inductors with low values of $\ell$ to compact kinetic inductors with high values of the sheet and linear inductance. For instance, in the SFQ5ee node we make inductors for biasing ERSFQ circuits by using a 40-nm thick MoN$_x$ layer with kinetic inductance of 8 pH/sq. The third objective of this work is to increase the scale of integration by developing a fabrication process to implement kinetic inductors in SFQ cells.

## II. PROCESS LINEWIDTH REDUCTION TO 250 nm

### A. Resistors

The typical spacing between shunt resistors in SFQ circuits is a few times larger than the resistor linewidth. We used 248-nm photolithography and high density plasma etching in Cl$_2$-based chemistry to pattern Ti/MoN$_x$/Ti resistor layer into meander-type structures (a snake in a comb) with 200 and 250 nm linewidth and 700 nm spacing between the lines. The snake length was slightly over 3 cm. Titanium was used as an adhesion layer; MoN$_x$ thickness was 40 nm. The structures were used to check for shorts and measure the linewidth and resistance uniformity. The percentage of structures without open lines and shorts between adjacent lines was 100%. The typical on-wafer and wafer-to-wafer variation (standard deviation σ) of the resistance of MoN$_x$ resistors obtained at different widths are given in the first two rows of Table I.

TABLE I. RESISTANCE VARIATION OF Ti/MoN$_x$/Ti AND Nb MEANDERS

| Linewidth, $w$ (nm) | Spacing, $s$ (nm) | On-wafer resistance variation, 1σ (%) | Wafer-to-wafer variation, 1σ (%) |
|---|---|---|---|
| 200 | 700 | 3.2 | 1.4 |
| 250 | 700 | 2.7 | 2.6 |
| 250 (Nb) | 250 | 3.9 | 4.2 |
| 250 (Nb) | 350 | 4.0 | 4.9 |

The typical scanning electron microscope (SEM) images of a patterned resistor structure with 250 nm linewidth and 250 nm spacing is shown in Fig. 2a.

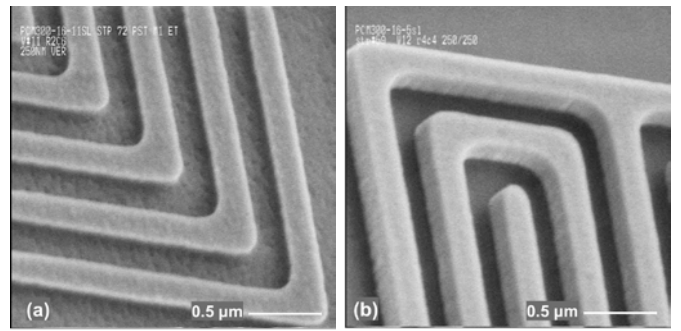

Fig. 2. SEM images of the etched features with 250-nm width and 250-nm spacing: (a) Ti/MoN$_x$/Ti resistors; (b) 200-nm thick Nb inductors.

The on-wafer uniformity of resistors was found to be primarily determined by the uniformity of the sheet resistance of Ti/MoN$_x$/Ti films on 200-mm wafers. Therefore, reproducibility of the mean resistance from wafer-to-wafer was typically better than within the wafer variation.

### B. Nb Inductors

The typical images of 250-nm-wide features patterned in a 200-nm-thick Nb inductor layer are shown in Fig. 2b. The linewidth uniformity and wafer-to-wafer repeatability were characterized by SEM and room temperature resistance measurements. The latter results are given in the two bottom rows of Table I, corresponding to Nb line spacing of 250 nm and 350 nm. The test structure resistance variation was somewhat larger for Nb inductors than for MoN$_x$ resistors, most likely because the Nb film thickness is 5x larger.

## III. INDUCTANCE AND MUTUAL INDUCTANCE OF DEEP-SUBMICROMETER INDUCTORS

Parasitic coupling of closely spaced inductors, i.e., their mutual inductance, is another factor limiting the minimum spacing and the scale of integration. On the other hand, large mutual inductance is important for making compact transformers for circuits with ac biasing, e.g., based on Quantum Flux Parametron (QFP) [8].

We measured the self-inductance and mutual inductance of Nb striplines and microstrips of various widths, $w$, at different spacing, $s$, of the parallel signal strips, using a differential SQUID circuit described in [7]. The typical configurations studied are shown as insets in Fig. 3. The notations correspond to our SFQ4ee [1] and SFQ5ee [2] processes; the thicknesses of all Nb and dielectric layers are given in [2],[7]. The length of the striplines $l$ was 30 μm, much larger than $w$ and $s$.

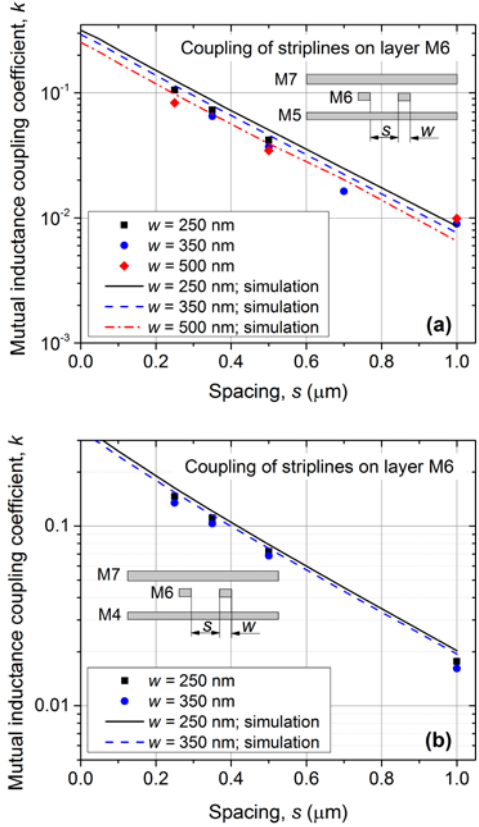

Fig. 3. Mutual inductance of two Nb striplines with width $w$ and spacing $s$ formed on the wiring layer M6 sandwiched between two ground planes: (a) M5 and M7 ground planes; (b) M4 and M7 ground planes. Symbols show the measurement results for various linewidths; lines show the simulation results using a 3D inductance extraction software wxLL [9].

The mutual inductance $M_{ij}$ between inductors $L_i$ and $L_j$ is defined as $\Phi_i = M_{ij}I_j$, where $\Phi_i$ is the magnetic flux induced in the loop $L_i$ by the current $I_j$ flowing in the loop $L_j$, where the self-inductance is defined by $\Phi_j = L_j I_j$. The mutual inductance can be expressed as $M_{ij} = k(L_i L_j)^{1/2}$, where $k$ is the coupling coefficient. For two identical inductors with $L_i = L_j = L$, the mutual inductance is simply $M_{ij} = kL$.

Fig. 3 shows the mutual inductance coupling coefficient $k$ for the two types of stripline inductors as a function of their spacing. The coupling coefficient weakly depends on the width of the striplines and exponentially depends on their spacing, reaching about 0.3 at zero spacing. For spacings below about 350 nm, the mutual inductance of the stripline inductors exceeds 10% of their self-inductance. This large mutual inductance should be explicitly taken into account in designing dense superconducting circuits.

However, for making compact transformers, high mutual inductance is desirable. Fig. 4 shows the mutual inductance of two 250-nm-wide microstrips formed on adjacent layers above the same ground plane as a function of their spacing.

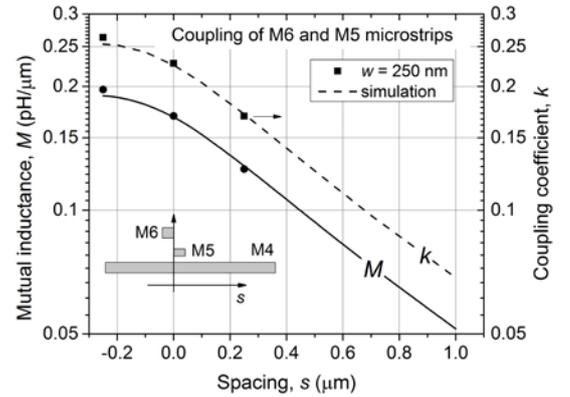

Fig. 4. Mutual inductance per unit length and the coupling coefficient of two microstrip inductors with width of 250 nm on two adjacent Nb layers M5 and M6 above the M4 ground plane as a function of spacing between their edges. Solid and dashed curves are simulations using a 3D inductance extractor.

ACKNOWLEDGMENT

We would like to thank M.A. Gouker, E. Dauler, and M.A. Manheimer for their interest in and support of this work; V.K. Semenov for the discussions, and M.M. Khapaev for access to his 3D inductance extractor.